\documentclass[10pt,twocolumn,letterpaper]{article}

\usepackage{cvpr}
\usepackage{graphicx}
\usepackage{amsmath}
\usepackage{amssymb}
\usepackage{booktabs}

\usepackage[pagebackref,breaklinks,colorlinks]{hyperref}

\usepackage[capitalize]{cleveref}
\crefname{section}{Sec.}{Secs.}
\Crefname{section}{Section}{Sections}
\Crefname{table}{Table}{Tables}
\crefname{table}{Tab.}{Tabs.}

\def\confName{CVPR}
\def\confYear{2022}

\begin{document}

\title{Exploring Detection-based Method For Speaker Diarization @ Ego4D Audio-only Diarization Challenge 2022}

\author{Jiahao Wang, Guo Chen, Yin-Dong Zheng, Tong Lu\\
\\
State Key Lab for Novel Software Technology, Nanjing University\\
{\tt\small wangjh@smail.nju.edu.cn, 
\{chenguo1177, ydzheng0331\}@gmail.com}\\
{\tt\small lutong@nju.edu.cn}
}
\maketitle

\begin{abstract}
   We provide the technical report for Ego4D audio-only diarization challenge in ECCV 2022. Speaker diarization takes the audio streams as input and outputs the homogeneous segments according to the speaker's identity. It aims to solve the problem of "Who spoke when." In this paper, we explore a Detection-based method to tackle the audio-only speaker diarization task. Our method first extracts audio features by audio backbone and then feeds the feature to a detection-generate network to get the speaker proposals. Finally, after postprocessing, we can get the diarization results. The validation dataset validates this method, and our method achieves 53.85 DER on the test dataset. These results rank 3$^{\text{rd}}$ on the leaderboard of Ego4D audio-only diarization challenge 2022.
\end{abstract}

\section{Introduction}
\label{sec:intro}

Speaker diarization is the task that processes input audio to the homogeneous segments according to the speaker's identity. It aims to answer the question "who spoke when" in a multi-speaker environment. It is widely applied in multi-speaker scenarios such as phone conversations, conferences, movies, and TV programs. 

To solve this challenging task, one major approach for speaker diarization is the clustering-based method, which clusters extracted audio features to get the segmentation for each speaker. Another representative work is Pyannote.audio\cite{pyannote}, which provides a set of trainable end-to-end neural building blocks that can be combined and jointly optimized to build speaker diarization pipelines. However, most clustering-based methods cannot solve speaker overlap as each time segmentation is allocated to one specific speaker.

In this report, we explore a detection-based speaker diarization method. In our pipeline, we first extract the audio feature by the audio-recognition backbone and then feed the feature to the advanced TAD detector model to generate dense proposals. Finally, after postprocessing, we can get the final results for audio-only diarization results.


\section{Approach}
We address audio-only diarization in three steps. First, we extract the audio feature by recognition backbone,  HuBERT\cite{hubert} approach excellent performance on speech recognition and other downstream tasks. To improve the performance of the proposal generator, we use HuBERT XLARGE to extract the audio feature. Then, we feed the feature to the proposal generator. To get precise results, we set the number of proposals to 1000, and after postprocessing, the final quantity of the proposals is 100.

\section{Experiments}
\subsection{Implemention Details}
We use HuBERT XLARGE to extract the audio feature, which is pre-trained 60,000 hours of unlabeled audio from the Libri-Light dataset and not fine-tuned. The sample rate of HuBERT is 16KHz, and the stride and window sizes are set to 320 and 400, respectively. After going through the audio backbone network, we can get a feature vector of 1280-dimension.

After feature extraction, we have tried two methods, one is Pyannote.audio's official model for speaker diarization, and another is the detection-based method. For Pyannote.audio model, we first use their default setting for inference and only get 89.74 DER on the Validation dataset. For the detection-based method, we use our detector to generate proposals for the advanced temporal action detection model ActionFormer\cite{actionformer} as our detector. We set the num\_classes as train dataset classes and feed the extracted feature to the detector. We adapt Soft-NMS\cite{softnms} to generate the final result for postprocessing. We got 66.54 DER on the validation dataset and 73.04 DER on the test dataset.

After trying the above two methods, we performed a brief data analysis on the training and test sets. For the training dataset, the number of speakers can reach 20, but only 4 in the validation dataset. Therefore, we set the maximum number of speakers in Pyannote.audio to 4 and the minimum number of speakers to 2, then re-inference. This time we got a DER of 62.89 on the validation dataset and 53.85 DER on the test dataset.

We also explore proposal fusion between Pyannote.audio and detection-based method. After generating the diarization result, we arrange the outputs of the two methods according to the scores and then select the top 100 results with the highest scores as the final output. We finally get 60.96 DER on the validation dataset. All results are shown in Table~\ref{result}.
\begin{table}
  \centering
  \begin{tabular}{c|cc}
    \toprule
    Method & Val DER & Test DER \\
    \midrule
    Pyannote(default) & 89.74 & - \\
    detection-based & 66.54 & 73.04  \\
    Pyannote + detection-based & \textbf{60.96} & -\\
    Pyannote(max\_speaker 4) & 62.89 & \textbf{53.85}\\
    \bottomrule
  \end{tabular}
  \caption{Results. The performance of Pyannote.audio, detection-based method, and score fusion.}
  \label{result}
\end{table}

\subsection{Limitation}

Although we reduced the DER to a certain value, in general, neither the detection-based method, Pyannote.audio nor score fusion performed well when the number of speakers was unknown. In addition, detection-based methods also poorly deal with features with overlapping speakers. Therefore, the next step will focus on making the model more accurate predictions in an environment where the number of speakers is unknown.

\section{Conclusion}

This paper explores a detection-based method to tackle audio-only speaker diarization tasks. Our method first extracts audio features by audio backbone and then feeds the feature to a detection-generate network to get the speaker proposals. Finally, after postprocessing, we can get the diarization results. The validation dataset validates this method, and our method achieves 53.85 DER on the test dataset.

{\small
\bibliographystyle{ieee_fullname}
\bibliography{egbib}

\begin{thebibliography}{1}\itemsep=-1pt

\bibitem{softnms}
Navaneeth Bodla, Bharat Singh, Rama Chellappa, and Larry~S Davis.
\newblock Soft-nms--improving object detection with one line of code.
\newblock pages 5561--5569, 2017.

\bibitem{pyannote}
Herv{\'e} {Bredin}, Ruiqing {Yin}, Juan~Manuel {Coria}, Gregory {Gelly}, Pavel
  {Korshunov}, Marvin {Lavechin}, Diego {Fustes}, Hadrien {Titeux}, Wassim
  {Bouaziz}, and Marie-Philippe {Gill}.
\newblock {pyannote.audio: neural building blocks for speaker diarization}.
\newblock 2020.

\bibitem{hubert}
Wei-Ning Hsu, Benjamin Bolte, Yao-Hung~Hubert Tsai, Kushal Lakhotia, Ruslan
  Salakhutdinov, and Abdelrahman Mohamed.
\newblock Hubert: Self-supervised speech representation learning by masked
  prediction of hidden units.
\newblock {\em IEEE/ACM Transactions on Audio, Speech, and Language
  Processing}, 29:3451--3460, 2021.

\bibitem{actionformer}
Chen{-}Lin Zhang, Jianxin Wu, and Yin Li.
\newblock Actionformer: Localizing moments of actions with transformers.
\newblock {\em CoRR}, abs/2202.07925, 2022.

\end{thebibliography}
}

\end{document}